\documentclass[11pt,article,single-spaced,amsmath,amssymb,superscriptaddress,nofootinbib,floatfix]{revtex4-2}
\usepackage{graphicx}
\usepackage{amsmath,amssymb}
\usepackage{bm}
\usepackage{mathrsfs}
\usepackage[hidelinks]{hyperref}
 
\begin{document}
\title{Spin-induced Lateral Acoustic Forces on Chiral and Achiral Particles}

\author{Yurou Jia}
\thanks{These authors contributed equally to this work.}
\affiliation{Department of Physics, City University of Hong Kong, Kowloon, Hong Kong, China}
\affiliation{Department of Physics and Center for Metamaterials Research,
Hong Kong University of Science and Technology, Hong Kong, China}

\author{Shanjun Liang}
\thanks{These authors contributed equally to this work.}
\affiliation{Division of Science, Engineering and Health Studies,
School of Professional Education and Executive Development,
Hong Kong Polytechnic University, Hong Kong, China}

\author{Wanyue Xiao}
\affiliation{Department of Physics, City University of Hong Kong, Kowloon, Hong Kong, China}

\author{Shubo Wang}
\email{shubwang@cityu.edu.hk}
\affiliation{Department of Physics, City University of Hong Kong, Kowloon, Hong Kong, China}

\begin{abstract}

Acoustic spin enables intriguing chiral sound-matter interactions, yet its mechanical consequences remain much
less explored. Here, we demonstrate spin-induced lateral acoustic forces on chiral and achiral particles
coupled to a substrate. Under plane-wave incidence, left- and right-handed chiral particles scatter
fields with opposite acoustic spin that selectively couple to counterpropagating guided waves, generating
lateral forces of opposite signs on the enantiomers. The lateral forces can also be realized for achiral
particles by employing acoustic vortex-beam illumination. We experimentally validate the spin-dependent
excitation of guided waves and lateral forces. Our results provide a route to spin-controlled acoustic manipulation and chirality-selective particle transport.

\end{abstract}
\maketitle

\noindent\textbf{Introduction}\\
Spin-dependent control of wave propagation and wave--matter interactions has emerged as an important theme in
modern wave physics~\cite{RodriguezFortuno2013NearField,Petersen2014ChiralWaveguide}. In optics, the interplay between spin angular momentum and directional momentum
transport underlies a variety of phenomena, including spin--orbit interactions~\cite{Bliokh2015SpinOrbit,VanMechelen2016SpinMomentum},
chirality-sorting optical forces~\cite{Wang2014LateralForce,Hayat2015ChiralitySorting,RodriguezFortuno2015LateralForces}, and geometric phases
~\cite{Bliokh2008Coriolis,Xiao2024TransverseSpinPB,Cheng2025RSPhase}. Although sound in fluids is
conventionally described by a scalar pressure field, its associated particle velocity field is intrinsically
vectorial and can exhibit local elliptical motion, thereby supporting nonzero acoustic spin
density~\cite{Bliokh2019SpinOrbitalAcoustic,Shi2019AcousticSpin,MuelasHurtado2022Polarization,Long2023GeometricRelations}. Acoustic spin underlies various phenomena~\cite{Burns2020AcousticSpin,Bliokh2024MomentumSpin,Gong2026AcousticOAM},
including geometric phases~\cite{Wang2018Topological,Xiao2026AcousticPB}, spin-orbit
interactions~\cite{Wang2021TransverseSound,Alhaitz2023SpinOrbital}, and circular dichroism
~\cite{Tong2023AcousticCD}. In particular, acoustic
spin--momentum locking has been demonstrated in guided and evanescent fields~\cite{Long2020AcousticSpinTransport,Weiner2022PseudoSpinHall}, where the sign
of the local transverse spin is locked with the wave propagation direction. These developments establish
acoustic spin as an additional degree of freedom for controlling directional energy and momentum transport,
raising a natural question of whether such spin-dependent transport can be converted into measurable mechanical
consequences. Acoustic forces provide a direct mechanical manifestation of wave--matter momentum
transfer
~\cite{Zhang2014Generation,Baresch2016,Zhang2018RadiationForce,Toftul2019AcousticForceTorque,Sepehrirahnama2022WillisCoupling,Smagin2024AcousticLateralRecoil,Toftul2026RadiationForcesReview}. Establishing a mechanism that converts acoustic spin into an imbalance of
transverse momentum transfer, and ultimately into a lateral mechanical force, can provide a distinct route
toward spin-controlled acoustic manipulation.

Here, we demonstrate lateral acoustic forces on chiral and achiral particles arising from acoustic
spin. For the chiral particles under plane-wave excitation, the scattered field is dominated by rotating
transverse dipoles that produce circularly-polarized velocity fields, which carry opposite transverse acoustic
spin for the two enantiomers. When coupled to a symmetry-preserving waveguide substrate, the
spin-polarized scattering near fields selectively excite counterpropagating guided waves through spin--momentum
locking, thereby inducing lateral acoustic forces of opposite signs on the enantiomers. We establish the
connections among local acoustic spin, multipolar scattering, and directional momentum transfer.
The same principle extends to achiral particles under acoustic-vortex-beam excitation. Our results
provide a route toward spin-controlled acoustic-force engineering and directional particle manipulation.\\

\noindent\textbf{Particle-substrate model system} \\
Figures~\ref{fig:1}(a) and \ref{fig:1}(c) schematically illustrate the proposed mechanism for generating spin-dependent lateral acoustic forces on chiral particles. An acoustic plane wave impinges obliquely on the particle at an incident angle \(\theta_{i}\) in  the $xz$-plane. The chiral particle generates a spin-polarized scattered field that couples to the waveguide substrate underneath and excites directional guided wave along the \(\pm y\) axis through spin--momentum locking. For the left-handed (LH) particle, the guided wave propagates along \(- y\), whereas for the right-handed (RH) particle it propagates along \(+ y\). The asymmetric momentum flux generates lateral radiation forces $F_y$ of opposite signs on the LH and RH particles.

\begin{figure}[ht]
\centering
\includegraphics[width=0.6\textwidth]{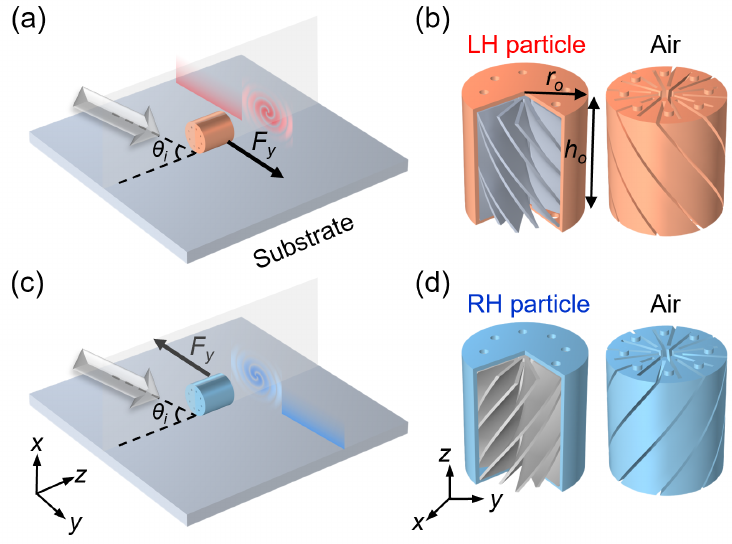}
\caption{Spin-induced lateral acoustic forces on chiral particles. (a) Oblique plane-wave illumination of a LH particle excites a directional guided wave and induces a lateral force $F_y$. (b) LH particle and its internal structures. (c)  Oblique plane-wave illumination of a RH particle excites a directional guided wave and induces a lateral force $-F_y$. (d) RH particle and its internal structures.}
\label{fig:1}
\end{figure}

Figures~\ref{fig:1}(b) and ~\ref{fig:1}(d) show the geometry of the chiral particles, which consist of helical internal channels coupled to the surrounding air through surface apertures. The LH and RH particles have identical geometric parameters but opposite structural handedness~\cite{SM}. For analytical description, we consider the chiral particle in $\mathbf r_0=(h,0,0)$, located above the upper surface of the substrate in $x=0$, under the illumination of a plane wave $p_{\mathrm{inc}}=e^{ik_0(-x\sin\theta_i+z\cos\theta_i)}$. The scattered pressure field can be expanded to $p=\sum_{\ell=0}^{1}\sum_{m=-\ell}^{\ell}a_{\ell m}h_\ell^{(1)}(k_0r')Y_{\ell m}(\hat{\mathbf r}')$, where $\mathbf r'=\mathbf r-\mathbf r_0$ and $r'=|\mathbf r'|$~\cite{Tsimokha2022multipolar}. In the subwavelength regime, the scattered field is mainly contributed by the monopole $Q$ and dipole $\mathbf D$ as $p^{\mathrm{sca}}=i\rho_0\omega[QG_0+\mathbf D\cdot\nabla' G_0]$, where $G_0=e^{ik_0r'}/(4\pi  r')$ is the free-space Green's function ~\cite{SM}.

The substrate modifies the response of the particles through multiple scattering ~\cite{Foldy1945MultipleScattering,Kleshchenko2024substrate}. Introducing the moment vector $\mathbf M=(Q,D_x,D_y,D_z)^{\mathsf T}$ and the background field $\mathbf B=(p_{\mathrm{bg}},\nabla p_{\mathrm{bg}})^{\mathsf T}_{\mathbf r_0}$, where $p_{\mathrm{bg}}$ comprises the incident field and its reflection from the substrate, the self-consistent response is governed by $\mathbf M=\boldsymbol\alpha_0(\mathbf B+\mathsf G_R\mathbf M)$, from which we obtain
\begin{equation}
\begin{aligned}
\mathbf M=\frac{\boldsymbol\alpha_0\mathbf B}{\mathbf I-\boldsymbol\alpha_0\mathsf G_R}
,
\end{aligned}
\label{eq:self-consistent}
\end{equation}
where $\boldsymbol\alpha_0$ is the bare polarizability matrix of the particle, and $\mathsf G_R$ accounts for the reflection of the substrate, 
\begin{equation}
\mathsf G_R=i\rho_0\omega
\begin{pmatrix}
G^R & (\nabla'G^R)^{\mathsf T}\\
\nabla G^R & \nabla(\nabla'G^R)^{\mathsf T}
\end{pmatrix},
\label{eq:green-matrix}
\end{equation}
where $\nabla$ and $\nabla'$ act on the observation and source coordinates, respectively. The reflected scalar Green function is expressed in the Weyl--Sommerfeld representation as~\cite{BrekhovskikhGodin1999,Quan2014impedance}
\begin{equation}
G^R(\mathbf r,\mathbf r_0)=\frac{i}{2}\int\frac{d^2\boldsymbol\kappa}{(2\pi)^2}\frac{R(\kappa,\omega)}{k_x}e^{i[\boldsymbol\kappa\cdot(\boldsymbol\rho-\boldsymbol\rho_0)+k_x(x+x_0)]}.
\label{eq:reflected-green}
\end{equation}
Here $\boldsymbol\rho=(y,z)$, $\boldsymbol\kappa=(\kappa_y,\kappa_z)$, $\kappa=|\boldsymbol\kappa|$, and $k_x=\sqrt{k_0^2-\kappa^2}$ is chosen on the outgoing or decaying branch. The function $R(\kappa,\omega)$ denotes the reflection coefficient of the substrate~\cite{SM}. The time-averaged radiation force can be expressed as~\cite{Toftul2019AcousticForceTorque,Sepehrirahnama2022WillisCoupling,Smagin2024AcousticLateralRecoil}
\begin{equation}
\begin{aligned}
F_i=\frac{1}{2\omega}\operatorname{Im}
\Big[Q^*\partial_i p+\sum_jD_j^*\partial_i\partial_j p \Big]
+\rho_0\Gamma_0\operatorname{Im}(D_i^*Q),
\end{aligned}
\label{eq:total-force}
\end{equation}
where $p=p_{\mathrm{bg}}+p_R$ is the total field acting on the particle, $p_R=i\rho_0\omega[QG^R+\mathbf D\cdot\nabla'G^R]$ is the substrate-reflected scattered field, and $\Gamma_0=k_0^3/(12\pi)$ is the direct-radiation recoil coefficient. All terms are evaluated using
the same substrate-dressed moments. For an in-plane isotropic substrate, the lateral radiation force along $y$ is~\cite{SM}:
\begin{equation}
\begin{aligned}
F_y=F_y^{\mathrm{bg}}+\rho_0\Big[
&(\Gamma_0+\operatorname{Im}I_t)
\operatorname{Im}(D_y^*Q)\\
&+\operatorname{Re}(I_{tx})
\operatorname{Im}(D_y^*D_x)
\Big],
\end{aligned}
\label{eq:lateral-force}
\end{equation}
where $F_y^{\mathrm{bg}}=(2\omega)^{-1}\operatorname{Im}[Q^*\partial_y p_{\mathrm{bg}}+\sum_jD_j^*\partial_y\partial_jp_{\mathrm{bg}}]$ denotes the background-field contribution, and the substrate-mediated reflection kernels are
\begin{equation}
\begin{aligned}
I_t&=\partial_y\partial_{y_0}G^R
=\frac{i}{8\pi}\int_0^\infty
\frac{\kappa^3R(\kappa,\omega)}{k_x}e^{ik_xs}d\kappa,\\
I_{tx}&=-i\partial_s I_t=\frac{i}{8\pi}\int_0^\infty
\kappa^3R(\kappa,\omega)
e^{ik_xs}\,d\kappa,
\end{aligned}
\label{eq:reflection-kernels}
\end{equation}
with $s=2h$. For the plane-wave excitation, $F_y^{\mathrm{bg}}=0$. The contribution proportional to $\operatorname{Im}(D_y^*Q)$ combines the monopole--dipole radiation recoil with its substrate-mediated counterpart; the term proportional to $\operatorname{Im}(D_y^*D_x)$ describes the substrate-mediated interference between two orthogonal dipoles. For a pair of enantiomers, the mirror transformation $y\rightarrow-y$ reverses both interference terms. Consequently, the two enantiomers experience lateral forces of opposite signs under the same incident plane wave, owing to their mirror-related scattering responses.\\

\begin{figure}[!t]
	\centering
	\includegraphics[width=0.6\textwidth]{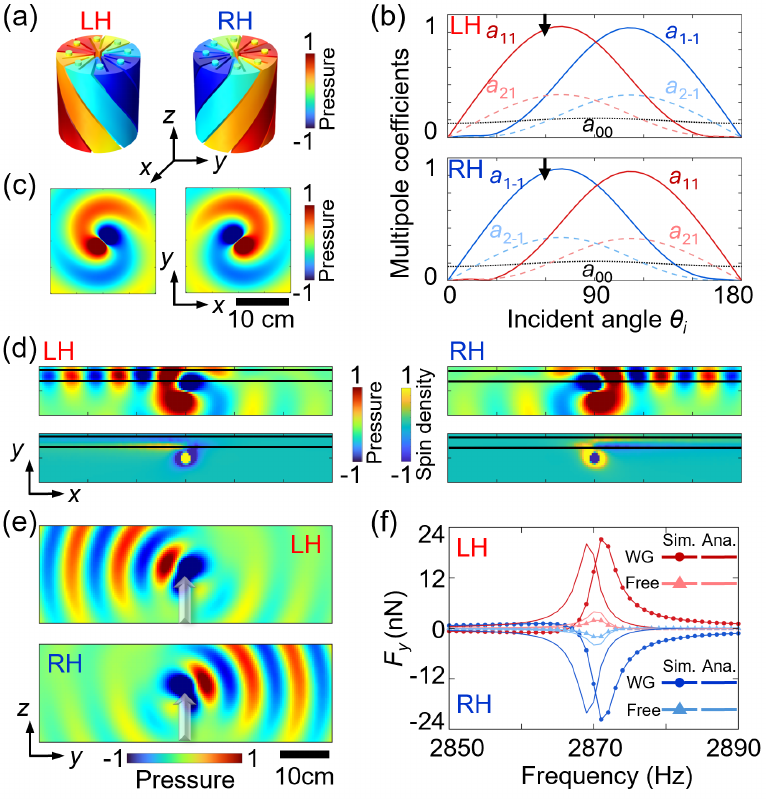}
	\caption{Multipolar scattering and lateral forces of chiral particles near a waveguide substrate. (a) Eigenmodes of the chiral particles. (b) Multipole amplitudes of LH (upper) and RH (lower) particles; arrows mark the selected incidence angle. (c) Free-space scattered fields. (d) Pressure (upper panels) and spin-density (lower panels) distributions for the LH (left) and RH (right) particles coupled to the substrate waveguide. (e) Directional guided pressure fields excited by the LH (upper) and RH (lower) particles. (f) Lateral force $F_y$ versus frequency with (WG) and without (Free) the waveguide: symbol lines denote full-wave simulation results, while solid lines denote analytical predictions. The substrate waveguide has a thickness of $20\,\mathrm{mm}$; its material density and compressibility are twice of that of background.}
	\label{fig:2}
\end{figure}

\noindent\textbf{Spin-dependent lateral force}\\ Figure~\ref{fig:2}(a) shows the eigenmode fields of the LH and RH particles at the dipole resonance frequency \(f=2871\,\mathrm{Hz}\). The LH particle supports a dipolar mode whose field rotates clockwise about the center axis, whereas the eigenfield of the RH particle rotates counterclockwise. To excite these modes, the particles (without the substrate) are illuminated by an external plane wave at the dipole-resonance frequency. Figure~\ref{fig:2}(b) shows the multipole expansion coefficients \(a_{lm}\) of their scattered fields under different incident angles \(\theta_{i}\). At \(\theta_{i} = 0^{\circ}\) and \(90^{\circ}\), the \(m = + 1\) and \(- 1\) components within each multipole order have equal amplitudes. At intermediate angles, however, \(a_{11}\) dominate over \(a_{1 - 1}\) for the LH particle, whereas the opposite hierarchy occurs for the RH particle. This reversal demonstrates handedness-dependent selectivity between opposite azimuthal multipole channels. Quadrupolar ($\ell=2$) coefficients are also shown to characterize higher-order scattering, whereas the analytical model retains $\ell\leq1$. At \(\theta_{i} = 60^{\circ}\), the scattered pressure fields exhibit opposite spiral wavefronts for the two enantiomers [Fig.~\ref{fig:2}(c)], corresponding to vortices carrying opposite orbital angular momentum. 

In the presence of the waveguide substrate, the fields of the particles couple asymmetrically to the waveguide: the LH (RH) particle couples predominantly to the left-(right-)propagating
guided mode, which is attributed to the spin-momentum locking of guided waves~\cite{Wei2020AcousticDirectionality, Long2020SymmetryDirectionality}. The near fields of the particles
and the guided waves carry a local spin density \(\mathbf{s} = \rho_{0}\operatorname{Im}\left( \mathbf{v}^{*}
\times \mathbf{v} \right)/(2\omega)\), where \(\mathbf{v}\) is the complex particle velocity field ~\cite{Shi2019AcousticSpin,Bliokh2019SpinOrbitalAcoustic}. The
enantiomers generate opposite spins \(s_{z}\) that selectively excite the
guided waves propagating in \(\pm y\) directions. The monopole generates an additional field 
propagating along \(z\) direction. Consequently, the total guided field propagates towards the upper-left
(upper-right) region of the substrate for the LH (RH) particle [Fig.~\ref{fig:2}(e)].

\begin{figure}[!t]
	\centering
	\includegraphics[width=0.6\textwidth]{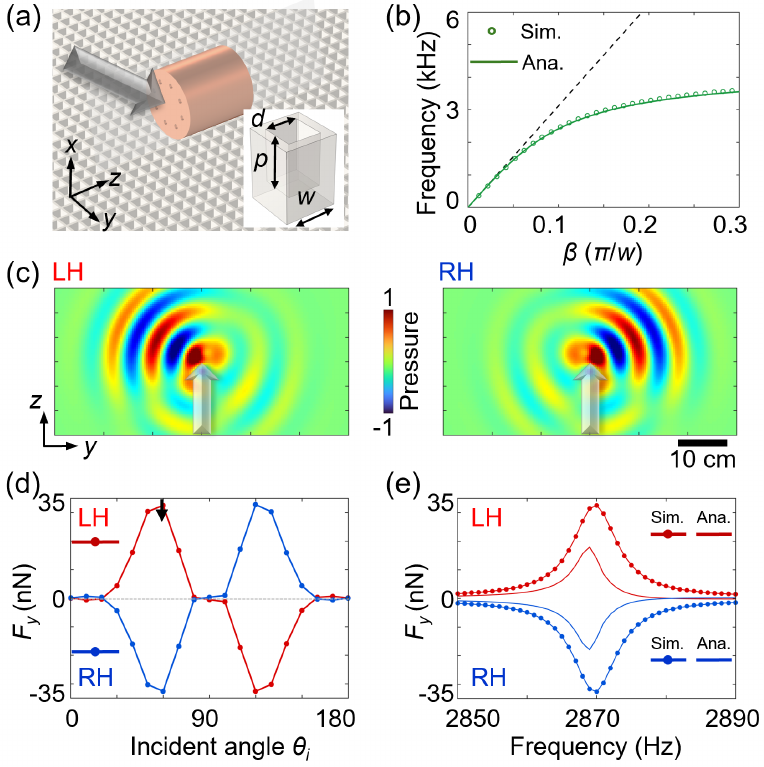}
	\caption{Lateral forces mediated by a perforated metamaterial substrate. (a) Schematic of a chiral particle placed above the substrate comprising a square array of rectangular cavities; the inset defines the lattice period $w$, cavity depth $p$, and opening width $d$. (b) Surface wave dispersion for the substrate with $p=20\,\mathrm{mm}$, comparing numerical simulations (symbols) with the effective-medium prediction (solid line); the dashed line is the free-space sound line. (c) Surface wave excited by the LH (left) and RH (right) particles. (d) Simulated lateral forces $F_y$ on the two enantiomers at different incidence angle $\theta_i$. (e) Lateral forces versus frequency, comparing simulation results (symbol lines) and analytical predictions (solid lines).}
	\label{fig:3}
\end{figure}

\begin{figure*}[!t]
	\includegraphics[width=1\linewidth]{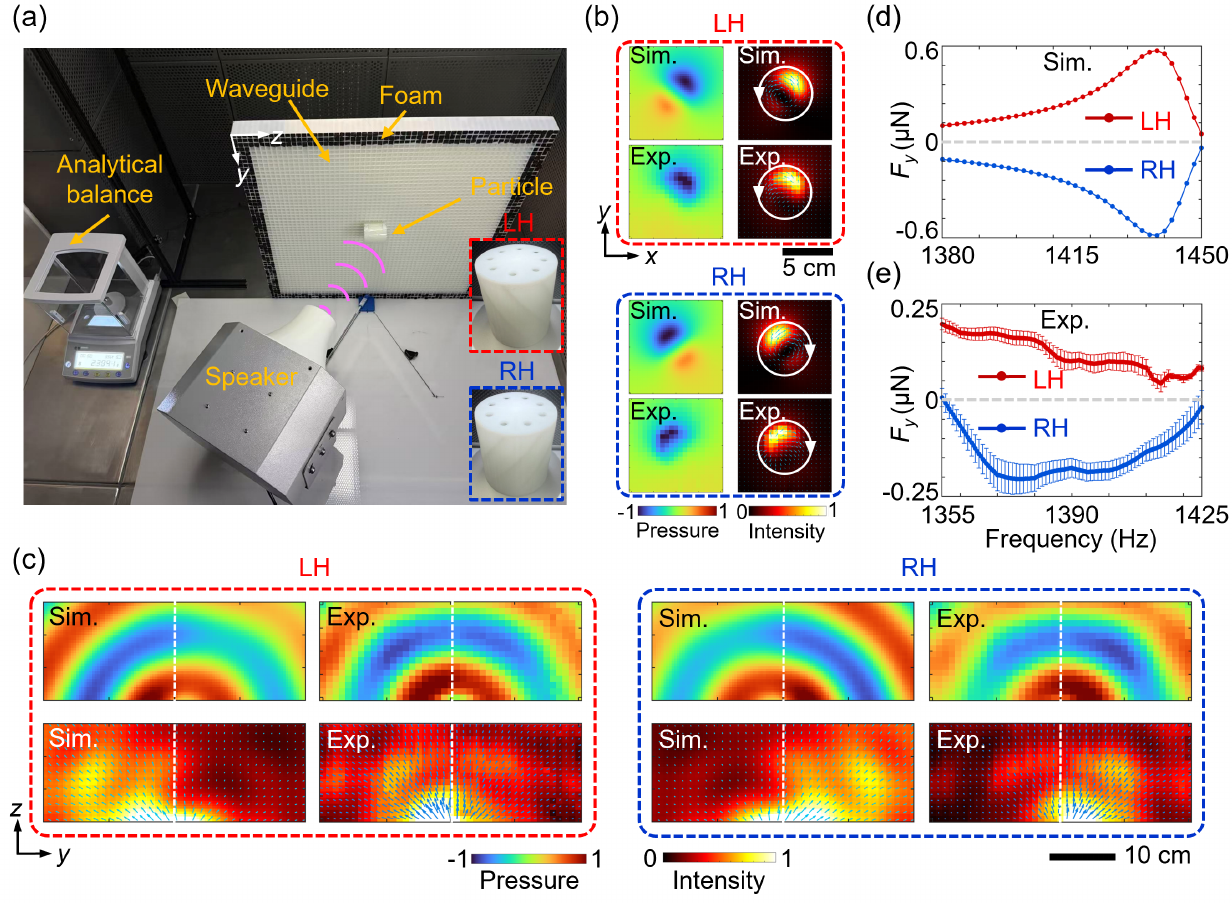}
	\caption{Experimental demonstration of chiral scattering and lateral acoustic forces. (a) Experimental setup and the fabricated LH and RH particles. (b) Simulated and measured free-space pressure profiles (left) and intensity patterns for the LH (upper group) and RH (lower group) particles. White arrows indicate the circulation direction of energy flux. (c) Simulated and measured pressure (upper) and intensity (lower) distributions above the substrate for the LH (left group) and RH (right group) particles. (d) Simulated lateral forces. (e) Experimentally measured lateral force spectra with error bars.}
	\label{fig:4}
\end{figure*}

The laterally asymmetric fields exert recoil forces of opposite signs on the enantiomers. Figure~\ref{fig:2}(f) shows that the particle-waveguide coupling enhances the peak lateral force from 1.87 nN (without substrate) to 21.2 nN (with substrate), corresponding to an enhancement factor of 11.3, with opposite force directions for the two enantiomers. The analytical predictions agree qualitatively with full-wave simulations, with minor resonance shifts arising from neglected higher-order multipoles. These results show that spin--momentum locking not only determines the direction of excited guided modes but also strongly enhances
the lateral acoustic force, providing a mechanism for chirality-selective particle manipulation.

To realize a practical waveguide in air, we employ a metamaterial substrate consisting of periodically arranged subwavelength cuboid holes [Fig.~\ref{fig:3}(a)]. The substrate supports surface waves with dispersion \(\beta = k_{0}\sqrt{1 + (d/w)^{4}\tan^{2}\left( k_{0}p \right)}\)~\cite{Xiao2026AcousticPB, Liu2018RainbowTrapping}, where $\beta$ is the propagation constant. Figure~\ref{fig:3}(b) shows good agreement between numerical simulations and the analytical effective-medium prediction, with the dispersion lying below the free-space sound line. The scattered fields in Fig.~\ref{fig:3}(c) exhibit opposite propagation directions for the LH and RH particles. Figures~\ref{fig:3}(d) and \ref{fig:3}(e) further show the corresponding lateral-force reversal versus incidence angle and frequency, with good agreement between full-wave simulations and analytical predictions.\\

\noindent\textbf{Experimental demonstration}\\
We experimentally validate the directional wave transport and the lateral forces. Figure~\ref{fig:4}(a) shows the experimental setup (see \cite{SM} for details), together with photographs of the fabricated LH and RH particles. Figure~\ref{fig:4}(b) compares the simulated and measured scattered fields of the LH and RH particles without the substrate, where the arrows indicate the energy circulation directions. The measured fields are in good agreement with the simulations; {the reduced visibility of the spiral wavefronts can be attributed to thermoviscous dissipation~\cite{SM}.}

Figure~\ref{fig:4}(c) shows the simulated and measured field distributions for the LH and RH
particles in presence of the metamaterial waveguide. For the LH (RH) particle, the acoustic field and energy flux are guided
predominantly toward the left (right). These
observations confirm the directional acoustic momentum transport induced by the coupling of the spin-polarized scattered field with the metamaterial substrate.

Figures~\ref{fig:4}(d) and \ref{fig:4}(e) show the simulated and measured lateral forces on the chiral particles above the substrate. The LH and RH particles experience lateral forces of opposite signs, in reasonable agreement with the simulations. The smaller measured forces may arise from dissipation, while the resonance frequency offset may be attributed to fabrication imperfections and uncertainties in particle positioning. These measurements confirm the spin--induced lateral forces in the chiral particle--waveguide system.\\

\noindent\textbf{Extension to achiral particles}\\ 
The mechanism can be extended to achiral particles, as schematically shown in Fig.~\ref{fig:5}(a). For an achiral particle illuminated by an acoustic vortex beam $p_{\mathrm{inc}}^{(\nu)}=e^{i\nu\varphi+ik_zz}$ with opposite topological charge $\nu=\pm1$ and $\varphi=\arg[(x-h)+iy]$, the near fields couple to counterpropagating guided waves through spin-momentum locking, giving rise to lateral forces of opposite signs.  
Figure~\ref{fig:5}(b) shows the geometry and eigenmodes of the achiral particle. Distinct from the chiral particle, the internal plates of the achiral particle are vertical rather than helical. The eigenmode exhibits a dipolar pressure distribution without chiral character. However, the acoustic spin of the incident vortex beams can excite chiral dipoles in the particle despite their achiral geometry, and the scattered pressure fields exhibit spiral patterns [Fig.~\ref{fig:5}(c)]. Coupling the near fields of the particle to the waveguide produces spin-dependent directional transport [Fig.~\ref{fig:5}(d)], consistent with the spin--momentum locking~\cite{BliokhNori2019Spin}.

\begin{figure}[ht]
\centering
\includegraphics[width=0.6\textwidth]{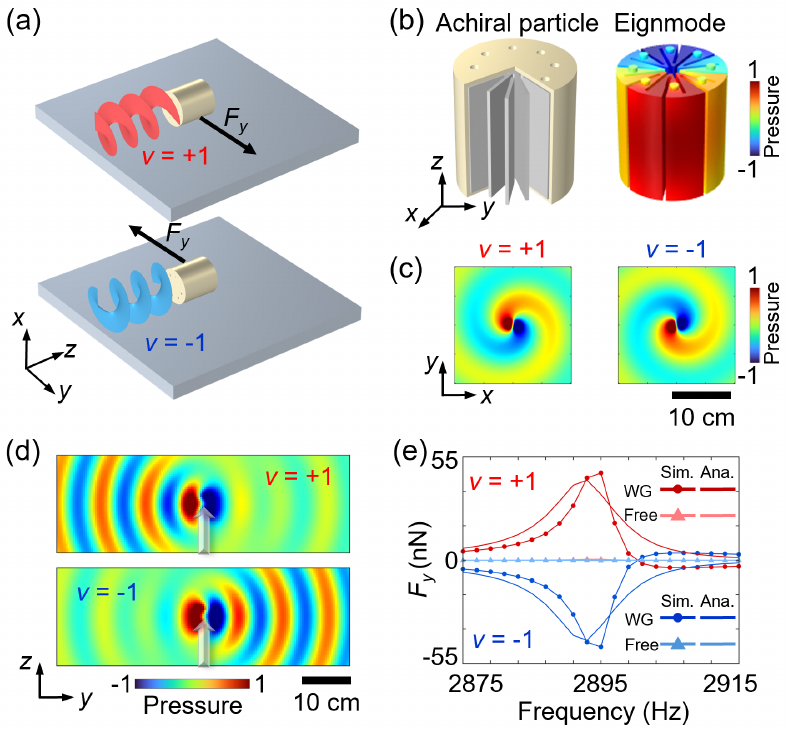}
\caption{Lateral acoustic forces on an achiral particle induced by an incident vortex beam. (a) Schematic of an achiral particle above a substrate waveguide, illuminated by vortex beams with opposite topological charges $\nu=\pm 1$, producing opposite lateral forces. (b) Geometry of the achiral particle and its eigenmode field. (c) Free-space scattered pressure fields for $\nu=+1$ and $\nu=-1$. (d) Scattered pressure fields coupled to the waveguide for $\nu=+1$ (upper) and $\nu=-1$ (lower). (e) Lateral forces with (WG) and without (Free) the waveguide. Symbols and solid lines represent full-wave simulations and analytical predictions, respectively. }
\label{fig:5}
\end{figure}

Figure~\ref{fig:5}(e) compares the simulated (symbol lines) and analytical (solid lines) lateral forces on the achiral particle with and without the waveguide. Unlike the plane-wave excitation case, the background-force term $F_y^{\mathrm{bg}}$ also contributes to the lateral forces. A sign change of the topological charge $\nu$ of the incident beam reverses the total lateral force, \(F_y^{(\nu)}=-F_y^{(-\nu)}\) [Fig.~\ref{fig:5}(e)].  Thus, tailoring the incident OAM so as to reverse the local acoustic spin can enable effective control of the lateral motion of achiral particles. Furthermore, when the substrate is removed, the lateral forces vanish due to the system's cylindrical symmetry.\\

\noindent\textbf{Conclusion} \\
In summary, we demonstrate spin-dependent lateral acoustic forces on chiral and achiral particles through spin–momentum locking. Chiral particles with opposite handedness generate scattered fields with opposite acoustic spin, which selectively excite counterpropagating guided modes and produce lateral recoil forces of opposite signs. We experimentally verify the directional wave transport and force reversal using a metamaterial substrate. We further extend this mechanism to achiral particles, where reversing the incident orbital angular momentum reverses the local acoustic spin and lateral force. Our findings provide a route to spin-controlled acoustic manipulation and chirality-selective particle transport.\\

\noindent\textbf{Acknowledgments} \\
The work described in this paper was supported by grants from the National Natural Science Foundation of China (No. 12322416) and Research Grants Council of the Hong Kong Special Administrative Region, China (Project No. AoE/P-502/20).\\

\noindent\textbf{References}\\
\vspace{-2.8cm}
\bibliography{references}
\clearpage

\clearpage
\begin{center}
    {\bfseries
    Supplemental Material for\\[0.1cm]
    ``Spin-induced Lateral Acoustic Forces on Chiral and Achiral Particles''
    }
\end{center}

	\title{Supplemental Material for \\
    ``Spin-induced Lateral Acoustic Forces on Chiral and Achiral Particles''}
\author{Yurou Jia}
\thanks{These authors contributed equally to this work.}
\affiliation{Department of Physics, City University of Hong Kong, Kowloon, Hong Kong, China}
\affiliation{Department of Physics and Center for Metamaterials Research,
Hong Kong University of Science and Technology, Hong Kong, China}

\author{Shanjun Liang}
\thanks{These authors contributed equally to this work.}
\affiliation{Division of Science, Engineering and Health Studies,
School of Professional Education and Executive Development,
Hong Kong Polytechnic University, Hong Kong, China}

\author{Wanyue Xiao}
\affiliation{Department of Physics, City University of Hong Kong, Kowloon, Hong Kong, China}

\author{Shubo Wang}
\email{shubwang@cityu.edu.hk}
\affiliation{Department of Physics, City University of Hong Kong, Kowloon, Hong Kong, China}

\maketitle

\setcounter{figure}{0}
\renewcommand{\thefigure}{S\arabic{figure}}
\setcounter{table}{0}
\renewcommand{\thetable}{S\arabic{table}}
\setcounter{equation}{0}
\renewcommand{\theequation}{S\arabic{equation}}

\begingroup
\makeatletter
\renewcommand*{\l@f@section}{%
    \addpenalty{\@secpenalty}%
    \addvspace{0.15em}%
}
\makeatother

\tableofcontents

\endgroup
\clearpage

\section{Multipole expansion}

The scattered acoustic pressure field of the chiral/achiral particles can be decomposed into a series of spherical harmonics (i.e., multipoles)~\cite{Tsimokha2022multipolar}:
\begin{equation}
	\label{SM_multipole}
	p^{\mathrm{sca}}(\mathbf{r})=
	\sum_{l=0}^{\infty}\sum_{m=-l}^{l}
	a_{lm} h_l^{(1)}(k_0r)Y_{lm}(\theta,\phi),
\end{equation}
where \(k_0\) is the acoustic wavenumber in the surrounding medium, \(a_{lm}\) denotes the complex scattering coefficient associated with the \((l,m)\) multipolar channel, \(h_l^{(1)}(k_0r)\) is the spherical Hankel function of the first kind describing the radial dependence of an outgoing spherical wave, and \(Y_{lm}(\theta,\phi)\) is the scalar spherical harmonic describing its angular dependence. The indices \(l\) and \(m\) denote the multipole order and azimuthal index, respectively. In particular, \(l=0\), \(1\), and \(2\) correspond to the monopole, dipole, and quadrupole contributions, respectively, while larger values of \(l\) represent higher-order multipolar components.

For an arbitrary acoustic scatterer, the multipole coefficients \(a_{lm}\) can be extracted by projecting the scattered pressure field onto the corresponding spherical-harmonic basis. Specifically, the coefficient associated with each multipolar channel is given by
\begin{equation}
	\label{SM_coefficient}
	a_{lm}=
	\frac{
		\displaystyle
		\int_{0}^{2\pi}\int_{0}^{\pi}
		p^{\mathrm{sca}}(r,\theta,\phi)
		Y_{lm}^{*}(\theta,\phi)
		\sin\theta\,d\theta\,d\phi
	}{
		\displaystyle
		h_l^{(1)}(k_0r)
		\int_{0}^{2\pi}\int_{0}^{\pi}
		Y_{lm}^{*}(\theta,\phi)
		Y_{lm}(\theta,\phi)
		\sin\theta\,d\theta\,d\phi
	}.
\end{equation}
Here, the superscript \(^*\) denotes complex conjugation. The magnitude \(|a_{lm}|\) quantifies the contribution of the corresponding multipolar channel to the scattered field, whereas the phase \(\phi_{lm}=\arg(a_{lm})\) characterizes its scattering phase. The multipole decomposition therefore provides a quantitative description of the relative contributions of different angular-momentum channels, enabling the identification of dominant scattering channels and their mutual interference.

\section{Acoustic radiation force calculation}
For a monochromatic acoustic field with angular frequency $\omega$, we adopt the harmonic convention $\widetilde{p}(\mathbf{r},t)=\mathrm{Re}\!\left[p(\mathbf{r})e^{-i\omega t}\right]$ and $\widetilde{\mathbf{v}}(\mathbf{r},t)=\mathrm{Re}\!\left[\mathbf{v}(\mathbf{r})e^{-i\omega t}\right]$, where $p$ and $\mathbf{v}$ are the total acoustic pressure and particle velocity fields, respectively, including both the incident and scattered contributions.  The surrounding fluid is assumed to be homogeneous, inviscid, and lossless, with mass density $\rho_0$ and sound speed $c_0$.  The acoustic radiation force (ARF) acting on the particle is calculated by integrating the time-averaged acoustic momentum flux density over a closed surface $S$ enclosing the particle
\begin{equation}
	\mathbf{F}
	=
	-\oint_S
	\langle\bm{\Pi}\rangle\cdot\mathbf{n}\,dS,
	\label{eq:arf_general}
\end{equation}
where $\mathbf{n}$ is the outward unit normal to $S$. For harmonic fields expressed in terms of peak complex amplitudes, the time-averaged momentum-flux tensor is~\cite{Zhang2014Generation}
\begin{equation}
	\langle\bm{\Pi}\rangle
	=
	\left(
	\frac{|p|^2}{4\rho_0c_0^2}
	-
	\frac{\rho_0|\mathbf{v}|^2}{4}
	\right)\mathbf{I}
	+
	\frac{\rho_0}{2}
	\mathrm{Re}
	\left(
	\mathbf{v}\otimes\mathbf{v}^{*}
	\right),
	\label{eq:arf_tensor}
\end{equation}
where $\mathbf{I}$ is the identity tensor, $\otimes$ denotes the dyadic product, and $^{*}$ denotes complex conjugation. Substituting Eq.~(\ref{eq:arf_tensor}) into Eq.~(\ref{eq:arf_general}) gives
\begin{equation}
	\mathbf{F}
	=
	\oint_S
	\left[
	\left(
	\frac{\rho_0|\mathbf{v}|^2}{4}
	-
	\frac{|p|^2}{4\rho_0c_0^2}
	\right)\mathbf{n}
	-
	\frac{\rho_0}{2}
	\mathrm{Re}
	\left\{
	\mathbf{v}^{*}
	\left(\mathbf{v}\cdot\mathbf{n}\right)
	\right\}
	\right]dS.
	\label{eq:arf_complex}
\end{equation}
Accordingly, the three Cartesian components of the ARF are given by
\begin{equation}
	F_x
	=
	\oint_S
	\left[
	\left(
	\frac{\rho_0|\mathbf{v}|^2}{4}
	-
	\frac{|p|^2}{4\rho_0c_0^2}
	\right)n_x
	-
	\frac{\rho_0}{2}
	\mathrm{Re}
	\left\{
	v_x^{*}
	\left(\mathbf{v}\cdot\mathbf{n}\right)
	\right\}
	\right]dS,
	\label{eq:Fx}
\end{equation}
\begin{equation}
	F_y
	=
	\oint_S
	\left[
	\left(
	\frac{\rho_0|\mathbf{v}|^2}{4}
	-
	\frac{|p|^2}{4\rho_0c_0^2}
	\right)n_y
	-
	\frac{\rho_0}{2}
	\mathrm{Re}
	\left\{
	v_y^{*}
	\left(\mathbf{v}\cdot\mathbf{n}\right)
	\right\}
	\right]dS,
	\label{eq:Fy}
\end{equation}
\begin{equation}
	F_z
	=
	\oint_S
	\left[
	\left(
	\frac{\rho_0|\mathbf{v}|^2}{4}
	-
	\frac{|p|^2}{4\rho_0c_0^2}
	\right)n_z
	-
	\frac{\rho_0}{2}
	\mathrm{Re}
	\left\{
	v_z^{*}
	\left(\mathbf{v}\cdot\mathbf{n}\right)
	\right\}
	\right]dS,
	\label{eq:Fz}
\end{equation}
where $\mathbf{v}\cdot\mathbf{n}=v_xn_x+v_yn_y+v_zn_z$ and $|\mathbf{v}|^2=|v_x|^2+|v_y|^2+|v_z|^2$.

\section{Analytical model for the lateral acoustic forces}
\label{sec:arf_theory}

We consider a particle at $\mathbf r_0=(h,0,0)$ in the homogeneous background fluid ($x>0$), with a planar substrate on the $x<0$ side. The substrate is assumed to be translationally invariant and isotropic in the $yz$ plane. The lateral force is along $y$ direction for the obliquely incident plane wave.

\subsection{Low-order multipole representation}

The outgoing scattered pressure field can be expanded as
\begin{equation}
p^{\mathrm{sca}}(\mathbf r)=
\sum_{\ell=0}^{1}\sum_{m=-\ell}^{\ell}
a_{\ell m}h_\ell^{(1)}(k_0r')Y_{\ell m}(\hat{\mathbf r}'),
\label{eq:S_scattered}
\end{equation}
where $\mathbf r'=\mathbf r-\mathbf r_0$ and $r'=|\mathbf r'|$. The coefficients for the monopole $\ell=0$ and dipole $\ell=1$ components are collected in
\begin{equation}
\mathbf A=(a_{00},a_{10},a_{1,+1},a_{1,-1})^{\mathsf T}.
\label{eq:S_four_channels}
\end{equation}
We introduce the generalized monopole--dipole source
\begin{equation}
\mathcal S(\mathbf r)=
Q\delta(\mathbf r-\mathbf r_0)
-\mathbf D\cdot\nabla\delta(\mathbf r-\mathbf r_0),
\label{eq:S_source}
\end{equation}
where $Q$ denotes the monopole, and $\mathbf D=(D_x,D_y,D_z)$ is the dipole. This source generates radiation fields through  the outgoing free-space Green function:
\begin{equation}
\begin{aligned}
p^{\mathrm{sca}}(\mathbf r)
=i\rho_0\omega\big[
QG_0(\mathbf r,\mathbf r_0)
+\mathbf D\cdot\nabla'G_0(\mathbf r,\mathbf r_0)
\big],
G_0(\mathbf r,\mathbf r_0)
=\frac{e^{ik_0|\mathbf r-\mathbf r_0|}}
{4\pi|\mathbf r-\mathbf r_0|},
\end{aligned}
\label{eq:S_point_field}
\end{equation}
where $\nabla'$ denotes differentiation with respect to the source position $\mathbf{r}_0$.
The far-field asymptotic relations $h_\ell^{(1)}(k_0r')\sim(-i)^{\ell+1}e^{ik_0r'}/(k_0r')$ and $\nabla'G_0\sim-ik_0\hat{\mathbf r}'G_0$ give
\begin{equation}
\begin{aligned}
p^{\mathrm{sca}}
\sim\frac{e^{ik_0r'}}{k_0r'}
\left[-ia_{00}Y_{00}
-\sum_{m=-1}^{1}a_{1m}Y_{1m}\right]
=\frac{\rho_0\omega e^{ik_0r'}}{4\pi r'}
\left[iQ+k_0\mathbf D\cdot\hat{\mathbf r}'\right].
\end{aligned}
\label{eq:S_far_match}
\end{equation}
Matching the terms of monopole and dipole on both sides gives
\begin{equation}
\begin{aligned}
Q&=-\frac{2\sqrt{\pi}}{\rho_0\omega k_0}\,a_{00},\\
D_x&=\frac{\sqrt{6\pi}}{\rho_0\omega k_0^2}
(a_{1,+1}-a_{1,-1}),\\
D_y&=\frac{i\sqrt{6\pi}}{\rho_0\omega k_0^2}
(a_{1,+1}+a_{1,-1}),\\
D_z&=-\frac{\sqrt{12\pi}}{\rho_0\omega k_0^2}\,a_{10}.
\end{aligned}
\label{eq:S_moment_mapping}
\end{equation}
These relations enable calculating the monople and dipole moments based on far-field expansions.

\subsection{Self-consistent particle--substrate dressing}

The particle response is modified by multiple scattering between the particle and the substrate. The total field acting on the particle consists of the externally applied background field and the substrate-reflected scattered field, $p=p_{\mathrm{bg}}+p_R$, where $p_{\mathrm{bg}}$ includes the incident field and its reflection from the substrate. Introducing the multipole moment vector $\mathbf M=(Q,D_x,D_y,D_z)^{\mathsf T}$ and the background excitation vector $\mathbf B=(p_{\mathrm{bg}},\nabla p_{\mathrm{bg}})^{\mathsf T}$, the self-consistent response is expressed as
\begin{equation}
\begin{aligned}
\mathbf M
=\boldsymbol\alpha_0
(\mathbf B+\mathsf G_R\mathbf M)
\Longrightarrow \mathbf{M}=\frac{\boldsymbol\alpha_0\mathbf B}{\mathbf I-\boldsymbol\alpha_0\mathsf G_R}
,
\end{aligned}
\label{eq:S_dressed_moments}
\end{equation}
where $\boldsymbol\alpha_0$ is the dynamic polarizability matrix of the bare particle. The substrate-mediated interaction matrix $\mathsf G_R$ is defined as
\begin{equation}
\mathsf G_R=i\rho_0\omega
\begin{pmatrix}
G^R & (\nabla'G^R)^{\mathsf T}\\
\nabla G^R & \nabla(\nabla'G^R)^{\mathsf T}
\end{pmatrix},
\label{eq:S_green_interaction}
\end{equation}
where $G^R(\mathbf r,\mathbf r_0)$ is the reflected scalar Green function, $\nabla$ and $\nabla'$ denote differentiation with respect to the observation and source coordinates, respectively, and all matrix elements are evaluated at the particle position. The corresponding reflected scattered field is 
\begin{equation}
p_R(\mathbf r)=i\rho_0\omega[QG^R(\mathbf r,\mathbf r_0)+\mathbf D\cdot\nabla'G^R(\mathbf r,\mathbf r_0)].
\end{equation}

\subsection{Reflected Green function and Sommerfeld kernels}

For an in-plane isotropic substrate, the reflected scalar Green function can be expressed through the Weyl--Sommerfeld representation as~\cite{BrekhovskikhGodin1999,Quan2014impedance}
\begin{equation}
G^R(\mathbf r,\mathbf r_0)=\frac{i}{2}\int\frac{d^2\boldsymbol\kappa}{(2\pi)^2}\frac{R(\kappa,\omega)}{k_x}e^{i[\boldsymbol\kappa\cdot(\boldsymbol\rho-\boldsymbol\rho_0)+k_x(x+x_0)]}.
\label{eq:S_GR}
\end{equation}
Here $\boldsymbol\rho=(y,z)$, $\boldsymbol\kappa=(\kappa_y,\kappa_z)$, $\kappa=|\boldsymbol\kappa|$, and $k_x=\sqrt{k_0^2-\kappa^2}$ is selected on the outgoing and decaying branch, with $\operatorname{Im}k_x\geq0$ and $\operatorname{Re}k_x\geq0$. The reflection coefficient $R(\kappa,\omega)$ incorporates multiple internal reflections in a finite slab.

To evaluate the lateral force, we first derive the relevant mixed derivatives of $G^R$. Setting $s=x+x_0$ and  $\boldsymbol\rho=\boldsymbol\rho_0$, the tangential dipole--dipole interaction kernel is
\begin{equation}
\begin{aligned}
I_t(s)
\equiv\left.
\partial_y\partial_{y_0}
G^R(\mathbf r,\mathbf r_0)
\right|_{\boldsymbol\rho=\boldsymbol\rho_0}
=\frac{i}{2}\int
\frac{d^2\boldsymbol\kappa}{(2\pi)^2}
\frac{\kappa_y^2}{k_x}
R(\kappa,\omega)e^{ik_xs}
=\frac{i}{8\pi}\int_0^\infty
\frac{\kappa^3}{k_x}
R(\kappa,\omega)e^{ik_xs}\,d\kappa.
\end{aligned}
\label{eq:S_It_derivation}
\end{equation}
The positive factor $\kappa_y^2$ follows from $(\partial_y\partial_{y_0})e^{i\kappa_y(y-y_0)}=(i\kappa_y)(-i\kappa_y)e^{i\kappa_y(y-y_0)}$. The final equality follows by setting $\kappa_y=\kappa\cos\phi$, using $d^2\boldsymbol\kappa=\kappa\,d\kappa\,d\phi$, and performing the azimuthal integral $\int_0^{2\pi}\cos^2\phi\,d\phi=\pi$. Differentiating $I_t$ with respect to the normal separation $s$ gives
\begin{equation}
\begin{aligned}
I_{tx}(s)
\equiv-i\partial_sI_t(s)
=\frac{i}{8\pi}\int_0^\infty
\kappa^3R(\kappa,\omega)e^{ik_xs}\,d\kappa.
\end{aligned}
\label{eq:S_reflection_kernels}
\end{equation}
At the particle position, $s=2h$. Both kernels contain propagating ($\kappa<k_0$) and evanescent ($\kappa>k_0$) contributions. Any guided modes supported by the substrate enter through the pole structure of $R(\kappa,\omega)$, with the appropriate causal or small-loss prescription.

For completeness, we define $g(s)=G^R|_{\boldsymbol\rho=\boldsymbol\rho_0}$, $g_s=\partial_sg$, and $g_{ss}=\partial_s^2g$. The reflected interaction matrix in the multipole basis then can be written as
\begin{equation}
\mathsf G_R=i\rho_0\omega
\begin{pmatrix}
g&g_s&0&0\\
g_s&g_{ss}&0&0\\
0&0&I_t&0\\
0&0&0&I_t
\end{pmatrix}_{s=2h}.
\label{eq:S_green_matrix}
\end{equation}
This structure follows from planar isotropy and azimuthal parity. In particular, $\partial_yG^R=\partial_{y_0}G^R=0$, $\partial_xG^R=\partial_{x_0}G^R=g_s$, $\partial_x\partial_{x_0}G^R=g_{ss}$, and $\partial_y\partial_{y_0}G^R=\partial_z\partial_{z_0}G^R=I_t$. The vanishing entries of $\mathsf G_R$ do not imply a diagonal particle polarizability.

\subsection{Force induced by the total field}

The total field acting on the particle is $p=p_{\mathrm{bg}}+p_R$, excluding the particle's directly emitted field $p^{\mathrm{sca}}$. The recoil associated with the directly radiated scattered field is evaluated separately using the self-consistently dressed multipole moments.
For the generalized multipole moment in Eq.~(\ref{eq:S_source}), the force induced by the total field can be expressed as
\begin{equation}
\begin{aligned}
F_i
=\frac{1}{2\omega}\operatorname{Im}
\int d^3\mathbf r\,
\mathcal S^*(\mathbf r)\,
\partial_i p(\mathbf r)
=\frac{1}{2\omega}\operatorname{Im}
\left[
Q^*\partial_i p
+\sum_{j=x,y,z}
D_j^*\partial_i\partial_jp
\right]_{\mathbf r=\mathbf r_0}.
\end{aligned}
\label{eq:S_point_force}
\end{equation}
The second equality follows by integration by parts of the derivative of the delta distribution. The spatial derivatives act on the total pressure field, while the self-consistently dressed source moments are held fixed during differentiation. The lateral force contains two terms
\begin{equation}
\begin{aligned}
F_y^{\mathrm{bg}}
&=\frac{1}{2\omega}\operatorname{Im}
\left[
Q^*\partial_y p_{\mathrm{bg}}
+\sum_jD_j^*\partial_y\partial_jp_{\mathrm{bg}}
\right]_{\mathbf r_0},\\
F_y^R
&=\frac{1}{2\omega}\operatorname{Im}
\left[
Q^*\partial_y p_R
+\sum_jD_j^*\partial_y\partial_jp_R
\right]_{\mathbf r_0}.
\end{aligned}
\label{eq:S_background_return}
\end{equation}
The total force additionally contains the direct-radiation recoil of the directly emitted scattered field. Importantly, all contributions are evaluated using the same self-consistently dressed moments.

\subsection{Evaluation of the reflected lateral force}

At the position of the particle, planar isotropy gives $\partial_y^2G^R=-I_t$ and $\partial_y\partial_{y_0}G^R=I_t$. Furthermore, the relation $\partial_sI_t=iI_{tx}$ implies $\partial_y\partial_x\partial_{y_0}G^R=iI_{tx}$ and $\partial_y^2\partial_{x_0}G^R=-iI_{tx}$. The remaining mixed derivatives relevant to the lateral force vanish by azimuthal parity.
Applying these identities to Eq.~(\ref{eq:S_background_return}), we obtain
\begin{equation}
\begin{aligned}
\partial_y p_R
&=i\rho_0\omega D_yI_t,\\
\partial_y\partial_xp_R
&=-\rho_0\omega D_yI_{tx},\\
\partial_y^2p_R
&=-i\rho_0\omega QI_t
+\rho_0\omega D_xI_{tx},\\
\partial_y\partial_zp_R
&=0,
\end{aligned}
\label{eq:S_derivatives}
\end{equation}
where the derivatives are evaluated at $\mathbf r=\mathbf r_0$. The reflected contribution to the lateral force can therefore be written as
\begin{equation}
\begin{aligned}
F_y^R
=\frac{1}{2\omega}\operatorname{Im}\big[
Q^*p_{R,y}+D_x^*p_{R,yx}
+D_y^*p_{R,yy}+D_z^*p_{R,yz}
\big]_{\mathbf r_0}.
\end{aligned}
\label{eq:S_Fy_general}
\end{equation}
Substitution of Eq.~(\ref{eq:S_derivatives}) and collection of complex-conjugate pairs give
\begin{equation}
\begin{aligned}
F_y^R=\frac{\rho_0}{2}\operatorname{Im}\big[
iI_t(Q^*D_y-D_y^*Q)
+I_{tx}(D_y^*D_x-D_x^*D_y)
\big].
\end{aligned}
\label{eq:S_reflected_pairing}
\end{equation}
Using $Q^*D_y-D_y^*Q=-2i\operatorname{Im}(D_y^*Q)$ and $D_y^*D_x-D_x^*D_y=2i\operatorname{Im}(D_y^*D_x)$, this expression reduces to
\begin{equation}
F_y^R=\rho_0\left[
\operatorname{Im}(I_t)\operatorname{Im}(D_y^*Q)
+\operatorname{Re}(I_{tx})\operatorname{Im}(D_y^*D_x)
\right].
\label{eq:S_Fy_reflected}
\end{equation}
The first term originates from reflected monopole--dipole interference, whereas the second term arises from the substrate-mediated interference between orthogonal lateral dipole components.

\subsection{Direct-radiation recoil and the unified lateral-force formula}

The direct-radiation recoil is obtained from the momentum carried by the directly radiated scattered field. Using the far-field representation in Eq.~(\ref{eq:S_far_match}), the reaction force on the particle is
\begin{equation}
F_i^{\mathrm{rec}}=
-\frac{\rho_0\omega^2}{32\pi^2c_0^2}
\int_{4\pi}n_i
\left|iQ+k_0\mathbf D\cdot\mathbf n\right|^2d\Omega,
\label{eq:S_recoil_integral}
\end{equation}
where $\mathbf n$ denotes the radiation direction.
Expanding the modulus square yields
\begin{equation}
\begin{aligned}
\left|iQ+k_0\mathbf D\cdot\mathbf n\right|^2
={}|Q|^2+k_0^2|\mathbf D\cdot\mathbf n|^2
-2k_0\sum_j n_j\operatorname{Im}(D_j^*Q).
\end{aligned}
\label{eq:S_recoil_expansion}
\end{equation}
The monopole-only and dipole-only terms are even under $\mathbf n\rightarrow-\mathbf n$ and hence do not contribute to the net momentum flux. Only the monopole--dipole interference term contributes. Using $\int_{4\pi}n_in_j\,d\Omega=(4\pi/3)\delta_{ij}$ and $\omega=c_0k_0$, we obtain
\begin{equation}
F_i^{\mathrm{rec}}
=\rho_0\Gamma_0\operatorname{Im}(D_i^*Q),
\qquad
\Gamma_0=\frac{k_0^3}{12\pi}.
\label{eq:S_free}
\end{equation}
Here, $\Gamma_0$ is the direct-radiation recoil coefficient; the moments $Q$ and $\mathbf D$ are substrate-dressed moments. 

Combining the background contribution, the reflected-field contribution, and the direct-radiation recoil gives the complete monopole--dipole lateral force:
\begin{equation}
\begin{aligned}
F_y=F_y^{\mathrm{bg}}+\rho_0\Big[
(\Gamma_0+\operatorname{Im}I_t)
\operatorname{Im}(D_y^*Q)
+\operatorname{Re}(I_{tx})
\operatorname{Im}(D_y^*D_x)
\Big].
\end{aligned}
\label{eq:S_Fy_final}
\end{equation}
The total lateral force consists of four contributions: background-field force $F_y^{\mathrm{bg}}$, direct-radiation recoil $F_y^{\mathrm{rec}}=\rho_0\Gamma_0\operatorname{Im}(D_y^*Q)$, monopole--dipole interference force \(F_y^{\mathrm{MQ}}=\rho_0\operatorname{Im}(I_t)\operatorname{Im}(D_y^*Q)\), and dipole--dipole interference force $F_y^{\mathrm{DD}}=\rho_0\operatorname{Re}(I_{tx})\operatorname{Im}(D_y^*D_x)$. All force contributions are evaluated using the same self-consistently dressed monopole and dipole moments, which already incorporate multiple scattering between the particle and the substrate.

For plane-wave incidence in the $xz$ plane, both the incident and specularly reflected fields are independent of $y$. Therefore, $\partial_y p_{\mathrm{bg}}^{\mathrm{pw}}=\partial_y\partial_j p_{\mathrm{bg}}^{\mathrm{pw}}=0$, giving $F_y^{\mathrm{bg,pw}}=0$. Equation~(\ref{eq:S_Fy_final}) consequently reduces to
\begin{equation}
\begin{aligned}
F_y^{\mathrm{pw}}=\rho_0\Big[
(\Gamma_0+\operatorname{Im}I_t)
\operatorname{Im}(D_y^*Q)+\operatorname{Re}(I_{tx})
\operatorname{Im}(D_y^*D_x)
\Big].
\end{aligned}
\label{eq:S_Fy_plane}
\end{equation}
Thus, the lateral force arises from the direct-radiation monopole--dipole recoil and the substrate-mediated multipolar interference.

\begin{figure*}[t]
	\centering
	\includegraphics[width=0.8\columnwidth]{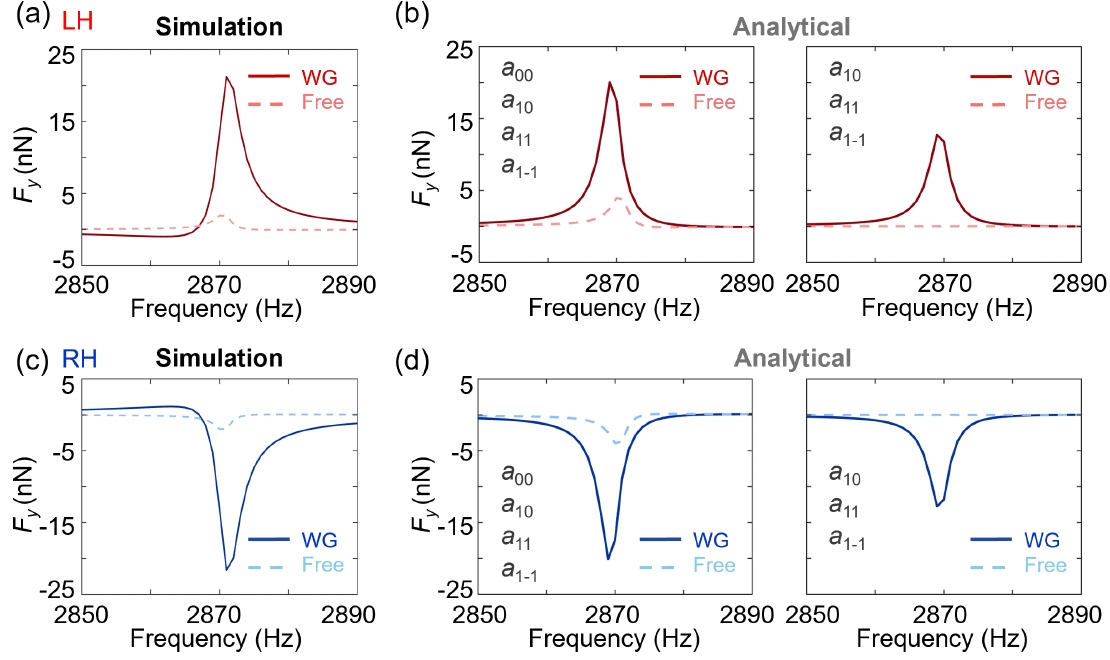}
	\caption{
		Full-wave and multipole analysis of the lateral ARF on the mirror-related LH and RH particles.
		(a) Full-wave lateral force on the LH particle with the waveguide (WG) and in free space (Free).
		(b) Corresponding self-consistent \(l\leq1\) multipole results. The left panel retains the four channels
		\(a_{00}\), \(a_{10}\), \(a_{11}\), and \(a_{1,-1}\), equivalent to the monopole \(Q\) and dipoles
		\(\mathbf D=(D_x,D_y,D_z)\), whereas the right panel shows the dipole-only result with the monopole channel suppressed.
		(c) and (d) Same as (a) and (b), respectively, for the RH particle.
		Solid and dashed curves denote the waveguide and free-space configurations, respectively.
	}
	\label{SM_analytical}
\end{figure*}

Figure~\ref{SM_analytical} further clarifies the physical origin of the
waveguide-enhanced lateral acoustic force.
As shown in Fig.~\ref{SM_analytical}(a), the simulated lateral force on the left-handed (LH) particle
is relatively weak in free space but is strongly enhanced in the presence of
the waveguide, indicating that the particle--waveguide interaction plays a
central role in the lateral recoil.
To identify the underlying multipolar mechanism, we construct an approximation
\(l\leq1\) model by retaining the lowest-order multipole channels
\(a_{00}\), \(a_{10}\), \(a_{11}\), and \(a_{1,-1}\), corresponding to the
monopole \(Q\) and three Cartesian dipoles
\(\mathbf D=(D_x,D_y,D_z)\).
The particle radius is \(12.5\,\mathrm{mm}\), as in the main text, and the
particle is separated from the waveguide by a gap of approximately
\(7\,\mathrm{mm}\).
The resulting lateral force, shown in the left panel of
Fig.~\ref{SM_analytical}(b), reproduces the main spectral features of the
full-wave calculation.
According to Eq.~(\ref{eq:S_Fy_final}), the force contains two distinct
contributions.
The term proportional to
\(\operatorname{Im}(D_y^*Q)\) originates from monopole--lateral-dipole
interference and gives the weak free-space recoil, together with its
reflection-induced correction \(\operatorname{Im}I_t\).
The second term,
\(
\rho_0\operatorname{Re}(I_{tx})
\operatorname{Im}(D_y^*D_x),
\)
arises from waveguide-mediated interference between the slab-normal dipole
\(D_x\) and the lateral dipole \(D_y\), and is strongly enhanced near the
waveguide resonance.
The small quantitative difference from the full-wave result is mainly
attributed to the neglected higher-order multipoles.

The role of the dipole--dipole interference is more evident in the right
panel of Fig.~\ref{SM_analytical}(b), where the monopole channel is
suppressed.
In this case, the free-space monopole--dipole recoil vanishes, whereas a
pronounced lateral force remains in the presence of the waveguide.
The force is therefore dominated by
\(
F_y
=
\rho_0\operatorname{Re}(I_{tx})
\operatorname{Im}(D_y^*D_x).
\)
This result explains why a strong lateral force can persist even when the
monopole response is weak, as in the circular-dipole-dominated regime
discussed in Fig.~5 of the main text.

Figures~\ref{SM_analytical}(c) and \ref{SM_analytical}(d) show the
corresponding results for the right-handed (RH) particle.
Because the LH and RH particles are related by mirror reflection, their
dipolar circularities have opposite signs, causing the dominant
waveguide-mediated lateral recoil to reverse direction.
These results identify a waveguide-mediated dipole--dipole recoil mechanism
that is distinct from the conventional free-space monopole--dipole
interference and provides the physical basis for the handedness-dependent
lateral force.

\begin{figure}[t]
	\centering
	\includegraphics[width=0.4\columnwidth]{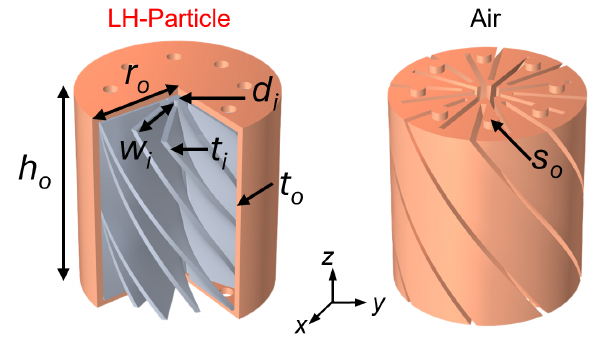}
	\caption{Schematic of LH particle and its internal structures.}
	\label{SM_particle}
\end{figure}

\section{Numerical simulations}
Numerical simulations were performed using COMSOL Multiphysics. The \textit{Pressure Acoustics, Eigenfrequency} interface was employed to calculate the eigenmodes of the particles, while the \textit{Pressure Acoustics} interface was used to obtain the acoustic pressure fields and spatial distributions. The surrounding medium was air with density \(\rho_0=1.21\,\mathrm{kg/m^3}\) and sound speed \(c_0=343\,\mathrm{m/s}\).

\noindent \textbf{Particle design.} The chiral particle was modeled as a cylindrical cavity with an outer radius of \(r_o=13.14\,\mathrm{mm}\), height of \(h_o=28.25\,\mathrm{mm}\), and shell thickness of \(t_o=1\,\mathrm{mm}\), as shown in Fig.~\ref{SM_particle}. Eight circular apertures with diameter \(s_o=2\,\mathrm{mm}\) and depth \(1\,\mathrm{mm}\) were symmetrically distributed on the top and bottom surfaces, enabling acoustic coupling between the external field and the internal cavity. Within the cavity, eight helical plates with width \(w_i=9\,\mathrm{mm}\) and thickness \(t_i=0.75\,\mathrm{mm}\) were arranged around an inner cylindrical rod with diameter \(d_i=3.28\,\mathrm{mm}\). An additional set of eight identical helical plates was distributed along the outer circumference. All helical plates possessed a pitch of \(P=105\,\mathrm{mm}\). The LH and RH particles shared the same geometric parameters and differed only in the handedness of the helical structures. An achiral particle was also considered, which maintained the same external dimensions and aperture configuration as the chiral particle, while its internal structure consisted of eight vertical plates along the inner circumference and eight additional identical plates along the outer circumference.

\noindent \textbf{Waveguide design.} Two types of waveguides were considered. For the idealized waveguide corresponding to Fig.~2(d,e), the chiral particle was placed above the waveguide surface with a vertical distance of \(7\,\mathrm{mm}\). The waveguide slab had dimensions of \(20\,\mathrm{mm}\) in thickness, \(600\,\mathrm{mm}\) in width, and \(600\,\mathrm{mm}\) in length. Its normalized mass density and compressibility were set to \(\bar{\rho}/\rho_0=2\) and \(\bar{\beta}/\beta_0=2\), respectively. In addition, a metamaterial waveguide was modeled as a rigid substrate containing periodically arranged rectangular holes in a square lattice. The unit cell had a period of \(w=5.5\,\mathrm{mm}\), a hole depth of \(p=20\,\mathrm{mm}\), and a hole width of \(d=4.5\,\mathrm{mm}\), all of which were deeply subwavelength compared with the operating wavelength.

\noindent \textbf{Multipole expansion and ARF calculation.} The chiral particle was first placed in free space, and a spherical integration surface enclosing the particle was employed to extract the multipole coefficients according to Eq.~\eqref{SM_multipole}. For the calculation of the ARF using Eq.~\eqref{eq:arf_complex}, the particle was positioned in close proximity to the metamaterial substrate. To accommodate the limited space between the particle and the substrate while ensuring that the integration surface fully enclosed the particle, an ellipsoidal integration surface was adopted. The calculated force is independent of the specific shape of the integration surface, provided that the surface is closed, fully encloses the particle, and lies entirely within the surrounding fluid domain, because the ARF is obtained from the momentum-flux integral over that surface. Both the chiral and achiral particles were modeled as acoustically rigid objects, corresponding to acoustic impedances much larger than that of air. The particles were illuminated by an obliquely incident plane wave with a pressure amplitude of \(1\,\mathrm{Pa}\) and an incidence angle of \(\theta_i\). Perfectly matched layers were applied at the outer boundaries of the simulation domain to suppress artificial reflections.

\noindent \textbf{Thermoviscous loss.} To account for thermoviscous dissipation, the \textit{Thermoviscous Acoustics} interface was coupled with the \textit{Pressure Acoustics} interface in the simulations. Thermoviscous effects were incorporated within the particle domain, including both viscous dissipation and thermal diffusion. The dynamic viscosity, bulk viscosity, heat capacity at constant pressure, and thermal conductivity of air were set to \(\mu=9.07\times10^{-6}\,\mathrm{Pa\cdot s}\), \(\mu_B=5.44\times10^{-6}\,\mathrm{Pa\cdot s}\), \(C_p=1005.4\,\mathrm{J/(kg\cdot K)}\), and \(\kappa=0.0258\,\mathrm{W/(m\cdot K)}\), respectively. No-slip and isothermal boundary conditions were imposed on all solid surfaces.

\begin{figure}[t]
	\centering
	\includegraphics[width=0.6\columnwidth]{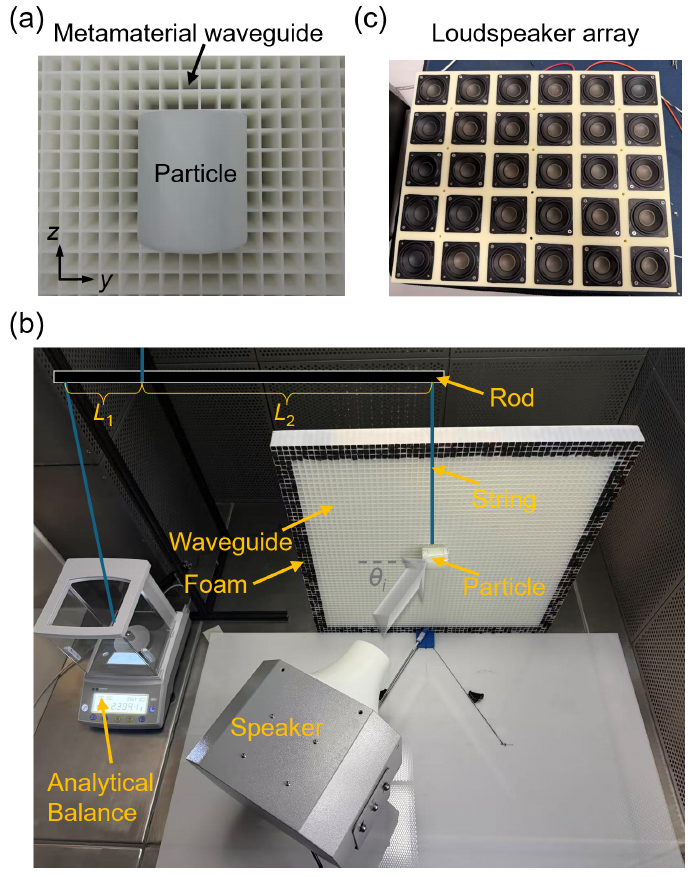}
	\caption{Experimental samples and measurement setups. (a) Photograph of the metamaterial substrate and the particle positioned above it. (b) Experimental setup used for the force and acoustic-field measurements. (c) Photograph of the loudspeaker array.}
	\label{SM_loud}
\end{figure}

\section{Experimental measurements}
\noindent \textbf{Sample preparation.}
To facilitate experimental fabrication and characterization, the LH and RH chiral particles were uniformly scaled up from the geometries used in the numerical simulations. The fabricated chiral particles had \(r_o=26.28\,\mathrm{mm}\), \(h_o=54.5\,\mathrm{mm}\), and \(t_o=1\,\mathrm{mm}\). Eight circular apertures, each with a diameter of \(d_o=4\,\mathrm{mm}\) and a depth of \(4\,\mathrm{mm}\), were symmetrically distributed over the top and bottom surfaces. Inside the cavity, eight helical plates with \(w_i=18\,\mathrm{mm}\) and \(t_i=1.5\,\mathrm{mm}\) were arranged around a central cylindrical rod with \(d_i=6.56\,\mathrm{mm}\). The helical pitch was \(P=210\,\mathrm{mm}\).

For the fabricated metamaterial waveguide, the corresponding dimensions were \(w=11\,\mathrm{mm}\), \(p=40\,\mathrm{mm}\), and \(d=9\,\mathrm{mm}\). Both the chiral particles and the metamaterial waveguide substrate were fabricated from photosensitive resin using a commercial 3D printer, as shown in Fig.~\ref{SM_loud}(a). Unless otherwise specified, all geometric dimensions of the experimentally fabricated structures were scaled by a factor of two relative to those used in the numerical simulations, while the shell thickness was kept unchanged.

\noindent \textbf{Force measurement.}
The force-measurement system was installed inside a sealed acoustic enclosure to minimize the influence of air convection and environmental disturbances, as shown in Fig.~\ref{SM_loud}(b). The inner walls of the enclosure were covered with sound-absorbing metamaterials to suppress unwanted acoustic reflections. To experimentally quantify the lateral ARF acting on the chiral particle, a high-power loudspeaker with an effective radiating aperture of \(10\,\mathrm{cm}\) in diameter was employed to generate an obliquely incident acoustic beam directed toward the metamaterial substrate. The loudspeaker was mounted on a movable platform equipped with rollers, allowing its position and orientation to be conveniently adjusted to control the propagation direction of the incident acoustic beam.

The chiral particle was suspended by an ultrathin string with a diameter of \(100\,\mu\mathrm{m}\), which was connected to one end of a rigid rod. The other end of the rod was mechanically coupled to a calibration mass placed on a digital analytical balance with a resolution of \(0.1\,\mathrm{mg}\). The lever-arm distances from the pivot point to the suspended particle and the calibration mass were denoted by \(L_2\) and \(L_1\), respectively, and were precisely adjusted prior to the measurements. This lever configuration converted and amplified the force-induced torque exerted by the lateral ARF, thereby enhancing the measurement sensitivity.

The lateral ARF was determined from the change in the balance reading upon application of the acoustic field. Specifically, the balance readings were recorded with the acoustic source switched off and on, and the corresponding ARF was calculated from the measured mass difference according to the calibrated lever-arm ratio \(L_1/L_2\). To reduce measurement uncertainty and suppress slow drifts in the balance reading, the acoustic source was periodically switched on and off at intervals of approximately \(2\,\mathrm{s}\). Multiple measurements were performed at each particle position and averaged to obtain the reported force values.

\noindent \textbf{Acoustic-field measurement.}
Acoustic-field measurements were performed in an anechoic chamber to minimize environmental noise and spurious reflections, as shown in Fig.~\ref{SM_loud}(b). The particle and metamaterial substrate were maintained in the same positions and configurations as in the force measurements. To generate an obliquely incident acoustic plane wave, the single loudspeaker used previously was replaced by a \(6\times5\) loudspeaker array, as shown in Fig.~\ref{SM_loud}(c), with the array positioned at the same nominal source location. The loudspeaker array was mounted on a rotatable mechanical stage, enabling precise control of the incidence angle, which was set to \(60^\circ\) for the measurements reported in the main text. The acoustic pressure field above the substrate was spatially scanned using a condenser microphone mounted on a computer-controlled translation stage. The measured signals were acquired using a digitizer and subsequently processed in MATLAB.

\begin{figure*}[t]
	\centering
	\includegraphics[width=0.8\columnwidth]{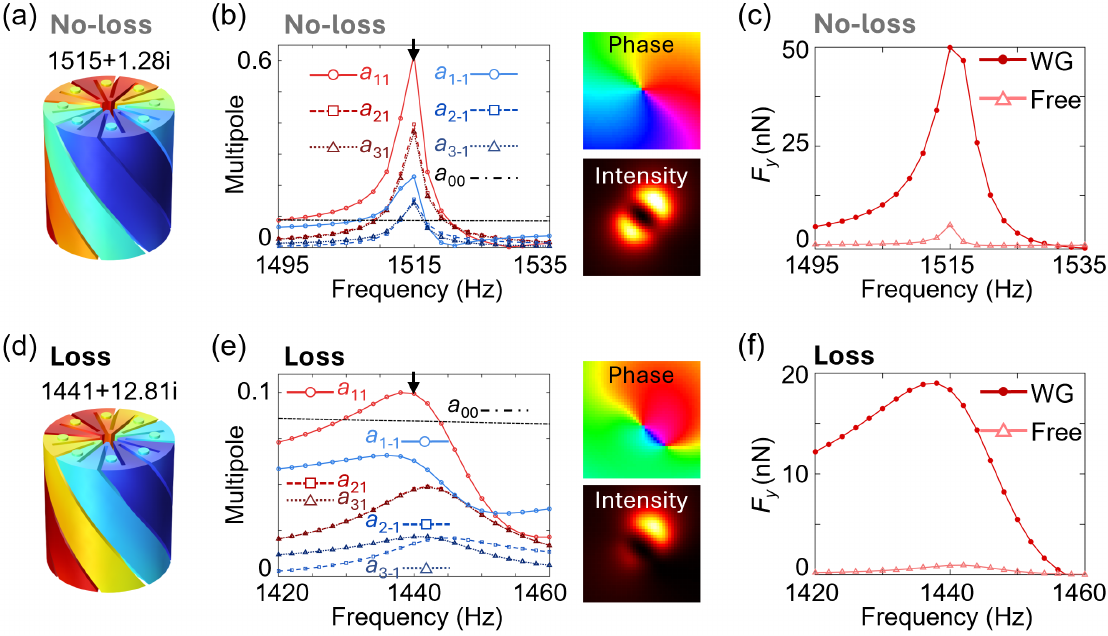}
	\caption{
		Influence of thermoviscous loss on the eigenmode, multipolar response,
		scattered field, and lateral ARF of the LH
		chiral particle.
		(a) Eigenmode of the LH particle in the absence of thermoviscous loss.
		The corresponding complex eigenfrequency is
		\(f_r+i f_i\), with \(f_r=1515\,\mathrm{Hz}\) and
		\(f_i=1.28\,\mathrm{Hz}\).
		(b) Multipole coefficients of the scattered field for an incidence angle
		of \(\theta_i=60^\circ\) in the lossless case. The insets show the phase
		and intensity distributions of the scattered field at
		\(1515\,\mathrm{Hz}\).
		(c) Frequency-dependent lateral ARF \(F_y\) acting on the LH particle
		in free space (Free) and above the metamaterial waveguide (WG) in the
		lossless case.
		(d)--(f) Same as (a)--(c), respectively, but with thermoviscous loss
		included. The complex eigenfrequency in (d) is characterized by
		\(f_r=1441\,\mathrm{Hz}\) and \(f_i=12.81\,\mathrm{Hz}\), and the
		scattered-field distributions in (e) are evaluated near
		\(1441\,\mathrm{Hz}\).
	}
	\label{SM_loss_sim}
\end{figure*}

\section{Influence of thermoviscous loss}
To assess the robustness of the proposed lateral-force mechanism under
realistic dissipative conditions, we further investigate the influence of
thermoviscous loss on the eigenmode, multipolar scattering response, scattered
field, and lateral ARF of the chiral particle. The
thermoviscous parameters used in the simulations are identical to those
specified in the section \textbf{Thermoviscous loss}, and the geometric
parameters are chosen to match those of the fabricated samples. Here, the LH
particle is considered as a representative example. The corresponding results
for the RH particle are not explicitly presented because two particles
are related by mirror symmetry.

We first consider the lossless case. As shown in
Fig.~\ref{SM_loss_sim}(a), the LH particle supports a chiral eigenmode with a
complex eigenfrequency \(f_r+i f_i\), where
\(f_r=1515\,\mathrm{Hz}\) and \(f_i=1.28\,\mathrm{Hz}\).
The eigenmode exhibits a pronounced rotational field pattern around the
\(z\) axis, consistent with the chiral resonance discussed in the main text.
To characterize the scattering response associated with this resonance, we
perform a multipole decomposition of the scattered field under oblique
incidence with \(\theta_i=60^\circ\). As shown in
Fig.~\ref{SM_loss_sim}(b), the positive-azimuthal-order components
\(a_{\ell1}\) are substantially stronger than their negative-order
counterparts \(a_{\ell-1}\) for \(\ell=1,2,3\), revealing a pronounced
imbalance between the two opposite angular-momentum channels. A finite
monopole component \(a_{00}\) is also present. At the resonance frequency
of \(1515\,\mathrm{Hz}\), the scattered-field phase exhibits a clear phase
winding, while the corresponding intensity distribution displays a
characteristic two-lobe pattern [insets of Fig.~\ref{SM_loss_sim}(b)].
These features are consistent with the dominance of the chiral scattering
channel near resonance.
Figure~\ref{SM_loss_sim}(c) compares the frequency-dependent lateral ARF
acting on the LH particle in free space and above the metamaterial waveguide.
In free space, only a relatively weak lateral force is observed. In contrast,
placing the particle above the waveguide produces a pronounced resonant
enhancement of \(F_y\). This enhancement arises from the coupling between the
chiral particle and the spin--momentum-locked guided mode, which produces an
asymmetric momentum exchange and consequently enhances the lateral momentum
transfer to the particle.

We next examine the effect of thermoviscous dissipation. As shown in
Fig.~\ref{SM_loss_sim}(d), after thermoviscous loss is included, the complex
eigenfrequency changes to \(f_r+i f_i\), with
\(f_r=1441\,\mathrm{Hz}\) and \(f_i=12.81\,\mathrm{Hz}\).
Compared with the lossless case, the resonance therefore shifts to a lower
frequency, while the substantially increased imaginary component indicates
stronger damping and a reduced quality factor. These changes originate from
viscous and thermal dissipation within the narrow acoustic channels of the
particle.
Thermoviscous loss also substantially modifies the multipolar composition of
the scattered field. As shown in Fig.~\ref{SM_loss_sim}(e), the imbalance
between \(a_{\ell 1}\) and \(a_{\ell -1}\) remains observable, indicating
that the handedness-dependent scattering response is preserved in the
presence of loss. However, the relative contribution of the monopole
component \(a_{00}\) becomes considerably more prominent than in the
lossless case. Correspondingly, the scattered field near
\(1441\,\mathrm{Hz}\) no longer exhibits the clear phase-winding structure
observed in the lossless case, and its intensity distribution is strongly
modified [insets of Fig.~\ref{SM_loss_sim}(e)]. This behavior can be
attributed to the loss-induced redistribution of the multipolar scattering
channels, particularly the increased relative contribution of the monopole
component.
The corresponding lateral-force spectra are shown in
Fig.~\ref{SM_loss_sim}(f). Thermoviscous dissipation broadens the resonant
response and reduces the peak lateral ARF, consistent with the increased
damping and reduced quality factor of the chiral resonance. Nevertheless,
the lateral force above the waveguide remains substantially stronger than
that in free space over the resonant frequency range. Therefore, although
thermoviscous loss weakens and broadens the resonance, the waveguide-mediated
enhancement of the lateral ARF persists. 

\begin{figure}[t]
	\centering
	\includegraphics[width=0.6\columnwidth]{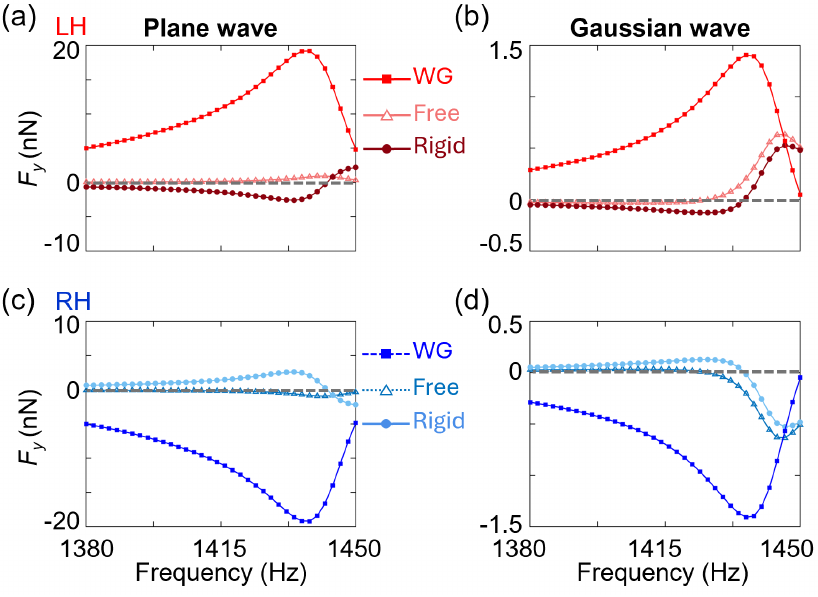}
	\caption{
		Effect of the incident beam profile on the lateral acoustic radiation
		force.
		(a) Frequency-dependent lateral ARF \(F_y\) acting on the LH particle
		under plane-wave excitation in free space (Free), above a rigid substrate
		(Rigid), and above the metamaterial waveguide (WG).
		(b) Same as (a), but under Gaussian-beam excitation with a beam width of
		\(10\,\mathrm{cm}\).
		(c),(d) Corresponding results for the RH particle under plane-wave and
		Gaussian-beam excitation, respectively. We assume equal intensity at the Gaussian beam center and of the plane wave.
	}
	\label{SM_Gauss}
\end{figure}

\section{Effect of the incident beam profile}
In the acoustic-field measurements, a loudspeaker array was employed to
generate an approximately plane acoustic wave. However, its acoustic output
was insufficient for a reliable measurement of the lateral acoustic radiation
force. For the force measurements, we therefore employed a high-power
loudspeaker with an effective radiating aperture of \(10\,\mathrm{cm}\) in
diameter. Owing to the finite aperture of the loudspeaker, the generated
acoustic field deviates from an ideal plane wave. To evaluate the influence of this finite beam width
on the measured lateral ARF, we compare the force responses under plane-wave
and Gaussian-beam excitation.

Figure~\ref{SM_Gauss}(a) shows the frequency-dependent lateral ARF \(F_y\)
acting on the LH particle under plane-wave excitation for three configurations:
the particle in free space, above a rigid substrate, and above
the metamaterial waveguide. Only a weak lateral force is observed in free
space and above the rigid substrate. In contrast, coupling the particle to the
metamaterial waveguide leads to a pronounced enhancement of \(F_y\) near the
resonance. This enhancement originates from the asymmetric momentum exchange
between the chiral particle and the spin--momentum-locked guided mode.

To account for the finite transverse extent of the experimentally generated
acoustic beam, we further consider a Gaussian beam with a beam width of
\(10\,\mathrm{cm}\), as shown in Fig.~\ref{SM_Gauss}(b). We notice that the the lateral forces in free space and above the rigid substrate are larger than those in the plane-wave case, which may be attributed to the enhanced direct radiation recoil.
Nevertheless, a pronounced resonant enhancement persists when the particle is
coupled to the metamaterial waveguide, and the resulting force remains
substantially larger than those in the two reference configurations.

Figures~\ref{SM_Gauss}(c) and \ref{SM_Gauss}(d) show the corresponding
results for the RH particle under plane-wave and Gaussian-beam excitation,
respectively. Reversing the handedness of the particle reverses the sign of
the waveguide-enhanced lateral force, whereas the overall spectral dependence
remains similar to that of the LH particle. Importantly, the pronounced
waveguide-mediated enhancement persists for both incident-beam profiles.
These results demonstrate that the finite beam width does not alter the essential mechanism of the
waveguide-enhanced lateral force.

\section{Lateral acoustic force for particles above a rigid substrate}

To further verify that the enhanced lateral ARF
originates from the coupling between the chiral particle and the metamaterial
waveguide, we compare the force responses above the metamaterial waveguide and a rigid substrate. A high-power loudspeaker with an effective
radiating aperture of \(10\,\mathrm{cm}\) in diameter was used as the acoustic
source, consistent with the configuration employed in the force measurements.
For a direct comparison between simulations and experiments, the force was
normalized with respect to the incident acoustic pressure and then rescaled
to an incident pressure amplitude of approximately \(20\,\mathrm{Pa}\).

\begin{figure}[tb]
	\centering
	\includegraphics[width=0.6\columnwidth]{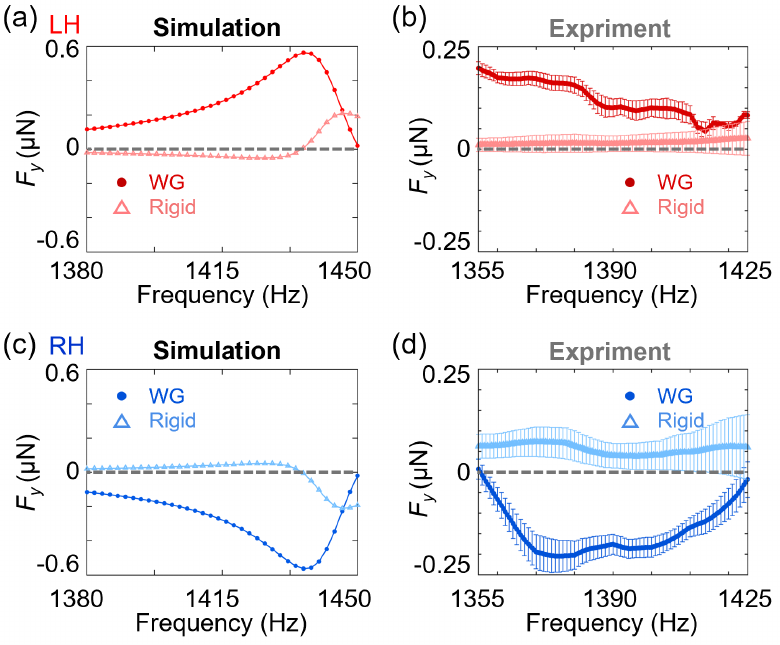}
	\caption{
		Comparison of the lateral ARF above the
		metamaterial waveguide and a rigid substrate.
		(a),(b) Simulated and measured frequency-dependent lateral ARF \(F_y\),
		respectively, for the LH particle.
		(c),(d) Corresponding simulated and measured results for the RH particle.
		Error bars in (a) and (c) indicate the experimental uncertainty obtained
		from repeated force measurements.
	}
	\label{SM_exprigid}
\end{figure}

Figures~\ref{SM_exprigid}(a) and \ref{SM_exprigid}(b) show the simulated and
measured lateral ARF \(F_y\), respectively, for the LH particle. In both
experiment and simulation, the force above the metamaterial waveguide is
substantially larger than that above the rigid substrate in the resonant
frequency range. The rigid substrate produces only a relatively weak lateral
force, indicating that the presence of a reflecting boundary alone is
insufficient to account for the pronounced force enhancement observed above
the metamaterial waveguide. In contrast, coupling of the chiral resonance to
the spin--momentum-locked guided mode enhances the asymmetric lateral momentum
exchange between the particle and the waveguide, resulting in a pronounced
positive \(F_y\).
The corresponding results for the RH particle are shown in
Figs.~\ref{SM_exprigid}(c) and \ref{SM_exprigid}(d). Reversing the particle
handedness reverses the sign of the waveguide-enhanced lateral force, while
the force above the rigid substrate remains comparatively weak. This
handedness-dependent force reversal is consistently reproduced in both the
measurements and simulations and agrees with the mirror-related responses of
the LH and RH particles.

The experimental force spectra exhibit broader features and smaller peak
magnitudes than the corresponding numerical results. These quantitative
differences may arise from thermoviscous dissipation, fabrication-induced
geometric deviations, the finite spatial profile of the incident acoustic
beam, and uncertainties associated with the particle location. Such
effects can also contribute to the frequency shift between the measured and
simulated force responses. Despite these quantitative differences, the
experiments reproduce the two key features predicted numerically: the
pronounced enhancement of the lateral ARF above the metamaterial waveguide
relative to the rigid substrate and the reversal of its direction upon
reversing the particle handedness. These results provide experimental
evidence that the enhanced lateral ARF is attributed to the chiral particle--waveguide coupling rather than the direct reflection of a rigid surface.

\end{document}